\documentclass[12pt,preprint]{aastex}

\def\msun{\>{\rm M_{\odot}}}

\shorttitle{Multiplicity Among Young Brown Dwarfs} 
\shortauthors{Ahmic et al.}

\begin{document}


\title{Multiplicity Among Young Brown Dwarfs and Very Low Mass Stars\altaffilmark{*}}

\author{Mirza Ahmic\footnote{e-mail: ahmic@astro.utoronto.ca}, Ray Jayawardhana, 
Alexis Brandeker\altaffilmark{**},\\ Alexander Scholz\altaffilmark{***}, Marten H. van Kerkwijk} 
\affil{Department of Astronomy \& Astrophysics, University of Toronto,
    Toronto, ON M5S 3H8, Canada}

\author{Eduardo Delgado-Donate} 
\affil{Instituto de Astrof\'isica de Canarias, E38205 - La Laguna, Tenerife, Spain}

\and 

\author{Dirk Froebrich}
\affil{Centre for Astrophysical and Planetary Science, University of Kent, Canterbury CT2 7NH, United Kingdom}

\altaffiltext{*}{We dedicate this paper to the memory of our co-author, Eduardo Delgado-Donate, who died in a hiking accident in Tenerife earlier this year.}
\altaffiltext{**}{Stockholm Observatory, AlbaNova University Center, SE-105 91 Stockholm, Sweden}
\altaffiltext{***}{School of Physics \& Astronomy, University of St. Andrews, North Haugh, St. Andrews, Fife KY16 9SS, United Kingdom}

\begin{abstract}
Characterizing multiplicity in the very low mass (VLM) domain 
is a topic of much current interest and fundamental importance. 
Here we report on a near-infrared adaptive optics imaging 
survey of 31 young brown dwarfs and VLM stars, 
28 of which are in the Chamaeleon I star-forming region, 
using the ESO Very Large Telescope. Our survey is sensitive 
enough to detect equal mass binaries down to separations of 
0.04--0.07$\arcsec$ ($\sim$6--10 AU at 160 pc) and, typically, 
companions with mass ratios ($q=m_2/m_1$) as low as 0.2 
outside of 0.2$\arcsec$ ($\sim$30 AU). We resolve the suspected 
0.16$\arcsec$ ($\sim$26 AU) binary ChaH$\alpha$ 2 and present two 
new binaries, Hn 13 and CHXR 15, with separations of 0.13$\arcsec$ 
($\sim$ 20 AU) and 0.30$\arcsec$ ($\sim$ 50 AU) respectively; 
the latter system is one of the widest VLM systems discovered to date. 
We do not find companions around the majority of our targets giving an 
overall binary frequency of $11^{+9}_{-6}\%$, thus confirming the trend 
for a lower binary frequency with decreasing mass. By combining our work 
with previous surveys of VLM objects (VLMOs) in other star forming regions, 
we arrive at the largest sample of young VLMOs (72) with high angular resolution 
imaging to date. Its multiplicity fraction is in statistical agreement 
with that for VLMOs in the field. In addition we 
note that many field stellar binaries with lower binding energies and/or wider 
cross sections have survived dynamical evolution and that statistical models 
suggest tidal disruption by passing stars is unlikely to affect the binary 
properties of our systems.  Thus, we argue that there is no significant 
evolution of multiplicity with age among brown dwarfs and VLM stars in OB and 
T associations between a few Myr to several Gyr. Instead, the observations to 
date suggest that VLM objects are either less likely to be born in fragile 
multiple systems than solar mass stars or such systems are disrupted very 
early (within the first couple of Myr). 
\end{abstract}

\keywords{stars: formation --  stars: low-mass, brown dwarfs 
-- stars: luminosity function, mass function -- binaries: general 
-- planetary systems}

\section{Introduction}
Over the past decade, hundreds of brown dwarfs (BDs) have been
identified in the solar neighborhood and in star-forming regions, yet there is no consensus on their origins. One possibility is 
that they form like stars, as a result of the 
turbulent fragmentation and collapse of molecular cloud cores 
(e.g., Padoan \& Nordlund 2004). Another scenario, which has gained 
popularity in recent years, is that BDs are stellar embryos ejected 
from multiple proto-stellar systems (Reipurth \& Clarke 2001; Bate et 
al. 2002). 

Recent observations have attempted to distinguish among these 
scenarios, by comparing physical properties of brown dwarfs and 
very low mass stars (which we refer to together as very low mass 
objects, or VLMOs, here) with those of solar-mass stars. Numerous 
studies have shown that many {\it young} VLMOs exhibit infrared 
excesses (e.g., Muench et al. 2001; Natta et al. 2002; Jayawardhana 
et al. 2003a; Scholz et al. 2007) and spectroscopic signatures 
of accretion (e.g., Jayawardhana et al. 2002, 2003b; 
Natta et al. 2004; Mohanty et al. 2005), indicative of disks. 
But both these diagnostics are only sensitive to the innermost 
portions of the disk and therefore cannot test whether BD disks 
are truncated, as expected in the ejection scenario. Recently, 
by modeling mid-infrared and millimeter emission, Scholz et al.
 (2006) found that $>$25\% of Taurus BDs in their sample have 
disks with radii $>$10 AU, contrary to predictions of certain 
ejection simulations, but some truncated disks may be hidden 
among their non-detections at 1.3mm. The spatial distribution 
and kinematics of BDs, in comparison with stars, do not provide 
definitive tests either; ejection models predict velocities 
comparable to the velocity dispersions of star-forming 
associations and clusters, i.e., 1-2 km/s (Moraux \& Clarke 2005). 

In this context, the multiplicity properties of VLMOs --such as 
frequency, separations, and mass ratios-- could be among the 
most important diagnostics of their origin. In particular, ejection 
models predict a very low rate of binaries among BDs ($<$5-8\%) and 
does not favor the formation of many wide ($>$20 AU) binary 
systems (e.g., Bate et al. 2002). Direct imaging surveys 
of (old) field VLMOs have yielded a binary frequency of $\sim$15\% for 
(apparent) separations $a>$1 AU, with maximum separations of 
$\approx$20 AU (e.g., Bouy et al. 2003; Burgasser et al. 2003). A 
search for spectroscopic binaries among the field VLMOs add another 
$\sim$11\% (Basri \& Reiners 2006). Open clusters older than $\sim$100 
Myr also appear 
to lack wide binary brown dwarfs (e.g., Mart\'in et al. 2003; Bouy 
et al. 2006a). At face value, these findings seem to support the 
ejection scenario --though typical binary separation may decrease 
at lower masses even in the turbulent fragmentation model. 

However, perhaps somewhat surprisingly, several wide binary VLM systems 
have been discovered within the past few years. Among them are: 
2MASS J1101-7732 ($a \approx$240 AU) in the $\sim$2-Myr-old Chamaeleon I 
star-forming region (Luhman 2004),  2MASS J1207-3932 ($a \approx$40 AU) 
in the $\sim$8-Myr-old TW Hydrae association (Chauvin et al. 2004; Mohanty 
et al. 2007), DENIS-P J161833-251750 and USCO-160028.4 ($a \approx$140 
and 120 AU, respectively) in the $\sim$5-Myr-old Upper Scorpius 
association (Luhman 2005; Bouy et al. 2006b), Oph 162225-240515 
($a \approx$240 AU) in 
the $\sim$1-Myr-old Ophiuchus region (Jayawardhana \& Ivanov 2006), and 
DENIS-J055146.0-443412.2 ($a>$200 AU) in the field (Bill\`eres et al. 2005). 
2MASS J1207-3932 and Oph 1622-2405 are especially intriguing because 
the secondaries have masses approaching the planetary regime. 
 
If wide binaries, such as these six, are common among VLMOs, that would 
present a critical challenge to ejection models. Recent high angular resolution surveys in Taurus and Upper Scorpious by Kraus et al. 
(2005, 2006) and Konopacky et al. (2007) have turned up 4 VLM binary systems (all in the Konopacky et al. 2007 sample) that would be considered wide by the ejection model standards. 
Here we report on an additional effort to constrain the frequency of wide VLM systems in other star forming regions. In particular, we describe a near-infrared adaptive optics imaging survey of 
28 VLMOs in Chamaeleon I designed to investigate whether multiplicity in the 
VLM regime depends on age. 

\section{Observations and Data Analysis}
Given its proximity (160\,pc; Whittet et al. 1997) and youth ($\sim$2\,Myr; 
Luhman 2004), the Chamaeleon I star-forming region is well suited for 
investigating multiplicity of young VLMOs. We observed 28 targets with spectral 
types of M5.25--M8 from the recent compilation by Luhman (2004); their membership 
in Cha I is supported by several diagnostics including proper motion, radial 
velocity and position on the H-R diagram. Six additional (random) VLM members 
reported in the census (3 M5.25, 1 M5.5, 1 M5.75, 1 M6) were not observed due 
to worsening atmospheric conditions. According to Baraffe et al. (1998, 2003) 
models, our targets span the mass range of $\sim 0.03\,-0.15 \msun$ (Table 1). 

We obtained near-infrared $H (1.66 \mu m)$- and $K_s (2.16 \mu m)$-band 
images of 26 of these targets with the 
NAOS-CONICA (NACO, for short) instrument on the 8.2m Yepun unit of the European 
Southern Observatory's Very Large Telescope on Paranal, Chile over three 
consecutive nights starting on March 24, 2006. We used the high-resolution lens 
of NACO, with a pixel scale of 13.26 mas/pix and a field-of-view of 
$13.6\arcsec\times13.6\arcsec$. For all observations, we utilized the infrared 
wavefront sensor (WFS) mode N90C10, which directs 90\% of the light to the WFS 
and the rest to CONICA. Targets themselves served as natural guide stars (NGSs) 
for the WFS. For each target, six 20 second exposures were combined to produce 
the final image. 

Two other Cha I targets, ChaH$\alpha2$ and ESO559, were observed in service 
mode between 2005 May--September using the same instrument set up, along with 
C-41 in Chamaeleon II (see Barrado y Navascu\'es \& Jayawardhana 2004), GY 5 
in $\rho$ Oph (Greene \& Young 1992) and 2MASSW J1207334-393254 in the TW Hydrae 
association (Gizis 2002). The latter is known to harbor a planetary mass companion 
at a separation of 
$\sim$40 AU (Chauvin et al. 2004; Mohanty et al. 2007), and thus served as a 
cross-check on our sensitivity. These five targets were observed only in the 
$K_s$-band, each with a total integration time of 18 minutes. 
We inspected the final images, reduced in a standard manner using the NACO 
pipeline software, visually to look for possible companions. In order to 
determine our ability to detect companions, we also derived contrast 
sensitivity curves through statistical analysis of the noise in annuli 
surrounding the central object. Then we calculated the 5-$\sigma$ detection 
limit as a function of separation from the primary for each of the targets. 
More details on the algorithm for deriving these sensitivities are given 
in Brandeker et al. (2006). We checked the accuracy of the procedure by 
artificially inserting faint companions into the images at various 
separations and trying to recover them; the results were consistent.  

We converted the contrast sensitivity limits into mass detection limits given 
the spectral type of the primaries, Luhman et al. (2003) temperature conversion scale and 
Baraffe et al. (1998, 2003) models for the appropriate age: 2 Myr for 
Cha I, 1 Myr for $\rho$ Oph and Cha II, and 8 Myr for TW Hydrae. 

To search for companions at smaller separations, we carried out 
point spread function (PSF) subtraction for the Cha I targets from the 
primary NACO run. To construct a template PSF, we averaged the 
PSFs of our brightest targets. By varying the combinations of objects used 
for the template PSF and comparing the results for consistency, we made sure 
not to include a binary in the template PSF. None of our objects shows a clear            
signature of a companion after PSF subtraction. For a more quantitative check, 
we used an approach similar to the procedure suggested by Bouy et al. (2006a): 
After PSF subtraction, we measured the residual scatter and looked for 
objects which show excessive noise in the PSF area. Again, none of our objects 
shows significantly increased noise, defined as 3$\sigma$ excess above the 
average. 

To probe the adequacy of this test, we inserted artificial companions for 
selected objects at separations of 10\,AU and 6\,AU with roughly 
equal brightness. Companions at 10\,AU were reliably detected, by visually 
inspecting the image before and after PSF subtraction. They 
also cause clearly enhanced scatter after PSF subtraction. Thus, we are      
confident that our companion search is complete down to 10\,AU for roughly 
equal mass systems. At 6\,AU, we can still recover artificial companions 
around the brightest of our targets, but for the fainter ones both the 
visual inspection and the noise analysis fail. We conclude that our 
companion search is probably not complete for separations $<10$\,AU. For 
the brightest $\sim 5$ objects, however, we can safely exclude the 
existence of roughly equal mass companions down to separations of 6\,AU.                 

\section{Results}
\subsection{Resolved Binaries}
We find three resolved binaries in the Cha I sample: ChaH$\alpha$~2, CHXR 15 and Hn~13 
(Fig. 1 and Table 2). The first was suspected to be a binary by Neuh\"auser 
et al. (2002) based on its elongated PSF in {\it Hubble Space Telescope} 
images. We are also able to recover the previously known companion to 
2M1207 in our images. We do not find companions to any other targets.  

Interestingly, ChaH$\alpha$~2, CHXR 15 and Hn~13 are all at the high mass end of our 
target sample: all have masses $>0.1\msun$, clearly above the sub-stellar 
boundary. Two of the binaries, ChaH$\alpha$~2 and Hn~13, have mass ratios close to unity 
($q=m_2/m_1>$0.9) and are slighthly wider ($20\mathrm{AU}<a<30\mathrm{AU}$) than most binaries 
found in the field. CHXR 15 stands out more: it has a lower mass ratio ($q=0.64$) 
and a wider separation ($\sim50\mathrm{AU}$). 

\subsection{Binary Frequency}
Fig. 2 shows our detection limits as a function of companion mass (according 
to Baraffe et al. models) and separation (assuming a distance of 160 pc). 
The survey is able to detect binaries with mass ratios ($q$) as low as 0.2 for all but one target 
at 50 AU, and planetary mass companions at larger separations. Our results 
suggest that wide VLM systems with ultra low mass companions, such as Oph 1622-2405 (Jayawardhana 
\& Ivanov 2006), are rare. We return to this point in the discussion.  

For comparison, we also plot the binaries found by Fischer \& Marcy (1992) 
among field stars with primary masses of 0.1-0.6\,$\msun$ and by Close et al. 
(2003), Burgasser et al. (2003) and Siegler et al. (2005) among field VLM objects with $0.05$--$0.1\msun$. Our survey would have 
detected nearly all binaries with $>$20 AU and $q>$ 0.6 in these comparison 
samples. For our Cha I sample, with primary masses between $\sim$0.02-0.2\,$\msun$, 
we find a binary frequency of $11^{+9}_{-6}\%$ in the regime where our survey is 
sensitive to companions ($a>20\mathrm{AU}, q>0.6$).  

\section{Discussion}
Our findings in Cha I are similar to what Kraus et al. (2005, 2006) have reported 
for VLMOs in Taurus and Upper Scorpius star-forming regions. A recent survey of Taurus by Konopacky et al. (2007) targeting mainly VLM stars has revealed a larger multiplicity fraction of wider systems with lower mass ratios, although the binary frequency is still in statistical agreement with our results\footnote{We are not 
able to extend the analysis to include the Bouy et al. (2006b) survey in Upper 
Scorpius because that paper does not list the targets and their properties.}. 
In fact, the four surveys target primaries in the same mass range and have 
comparable detection limits for companions\footnote{We removed three objects in the Konopacky et al. (2007) sample whose masses are greater than $0.15\msun$.}. Thus, we can combine the results 
of all four surveys (12 VLM targets in Upper Scorpius, 32 in Taurus and 28 in Chamaeleon I) to arrive at a more statistically significant sample of 
72 young VLMOs with an overall resolved binary frequency of $7^{+4}_{-3}\%$. If we just consider T-associations (Taurus and Chamaeleon I), then the multiplicity fraction is $5^{+5}_{-3}\%$. In both cases we use $a>8\mathrm{AU}$ and $q>0.8$ as our sensitivity limit. 

We can now compare the binary properties of young VLMOs with other populations 
of interest, such as field VLMOs and young stars, to explore possible variations 
with mass and/or age. We used a Fisher ``exact'' statistical test for comparing samples of binomial distributions, as detailed in Appendix B of Brandeker et al. (2006). The results are given in Table 3, where 
``likelihood'' refers to the probability that the multiplicity frequencies 
of the two samples are the same. 

\subsection{Mass Dependance of Binarity}
In order to explore the dependence of the binary frequency on mass we 
statistically compare several samples spanning different mass bins. 
As shown in Table 3 and Fig. 4, we constrain the $q$ and separation range as 
appropriate to ensure that the two samples being compared have the 
same sensitivity limits for companions. We ensure that samples that are compared are of similar age by only comparing binaries from young associations (Cha I, Taurus and Upper Scorpius). In addition we also address the potential concern about the environmental effects by performing every comparison with and without the 12 targets in Upper Scorpius as it is the only higher density association that contributes binaries to our combined sample. We find that our conclusions are not affected by inclusion of these additional binaries. 

We start by dividing the young VLMOs themselves into two groups, following 
Kraus et al. (2006): one with primary masses less than 0.07 $\msun$ and 
the other with primary masses greater than 0.07 $\msun$. All eight resolved 
VLMO binaries within the sensitivity limits (Fig. 4A) are in the higher mass bin. The likelihood of the two mass bins having the same binary frequnecy if only T-associations are compared is low ($\sim 8\%$), but not insignificant. If we increase the sample size by adding the Upper Scorpius members, 6 to the lower and 6 to the higher mass bin, the likelihood
that the two mass bins have the same binary distribution becomes even lower ($\sim 2\%$). In addition, we also note that 4 systems that are outside of the sensitivity range fall in the higher mass bin and thus further reinforce our point.
We can also compare 
the young VLMO sample with young stars ($\sim$0.57-1.40 $\msun$) in Taurus and Cha I 
surveyed by Ghez et al. (1993, 1997) in the sensitivity regime outlined in Fig. 4B. The result (likelihood 
of $\sim 1\%$ and $\sim 2\%$ if we include or exlude the Upper Scorpius sample respectively; Table 3) bolsters the case for a decreasing wide binary frequency with mass. 

What could account for this observed trend? One possibility is that less 
massive, and therefore fainter, primaries will have even fainter companions 
that might not be detectable by surveys. However, all of the samples presented here are sensitive to mass ratios of $q\geq0.5$ at 
separations greater than $30\mathrm{AU}$ and even lower mass ratios at wider separations (Fig 2. and Fig 4.) for the young 
VLMO sample, allowing us to rule out such a bias. Another possibility is 
that binary separations (i.e., system size) decreases with mass below our 
detection limits. A radial velocity survey of Cha I VLMOs finds only 
one candidate companion (Joergens 2006) at $\lesssim$0.1 AU among 
10 targets while Basri \& Reiners (2006) find a spectroscopic binary fraction of $\sim$11\% 
in the 0-6 AU separation range among field VLMOs. Imaging surveys of field 
VLMOs (eg., Close et al. 2003, Siegler et al. 2005), which have typical sensitivities of 
$q\sim0.85$ at $3\mathrm{AU}$, also do not reveal a significant number of companions closer 
to the primaries. Thus, it appears unlikely 
that a decrease in separations is the primary reason for the observed drop 
in multiplicity in the VLM regime. Instead, we conclude that the binary 
frequency most likely decreases with mass.

\subsection{Age Dependance of Binarity}
We investigate whether binarity in the VLM regime depends on age by 
comparing the young VLMO sample with the late M and early L field dwarfs 
surveyed by Close et al. (2003). They found three nearly equal mass 
$(q>0.85)$ pairs wider than $\sim$7.5 AU among 85 targets in the 
$\sim$0.05-0.1 $\msun$ mass range, resulting in a binary fraction 
of $4^{+6}_{-4}\%$, which is in good agreement with that for young VLMOs suggesting that binarity in the VLM regime does not depend strongly on age between a few Myr and several Gyr. One concern is that environments from which these two samples are drawn may be different. The young VLMO sample comes from low density T-associations (and OB-associations if we choose to include Upper Scorpius), while most of the field objects probably formed in higher density clusters. Thus one can argue that the multiplicity fraction we see today in the surveyed SFRs might decrease over time below that seen in the field. It is unlikely, however, that this will happen because the field VLMOs have survived more dynamically active environments.   

We also examine the stability of VLM binaries by calculating their binding 
energies in comparison to those of other binary populations. Fig.~3 shows 
that even though there is significant overlap of young and field VLM binaries 
on the binding energy vs. separation plot, young unbound associations also have 
a significant population of wider binaries. As this is the population that might 
be disrupted by dynamical encounters, exploration of stability of such systems 
is important. Regarding this problem, we note that a number 
of field star systems (those in the boxed area of the figure) are more 
fragile than the young VLM binaries, even though the former have evolved 
dynamically already. This is all the more surprising if one considers the fact 
that these field systems were most likely born in higher density regions where 
dynamical evolution would have had a more profound effect. Therefore, it is 
unlikely that the ``harder'' VLM 
binaries, would be disrupted, again supporting 
the conclusion that age is not a major factor affecting binarity in the 
VLM regime, at least when one considers OB and T associations\footnote{To 
date, 3 very wide ($\geq220$\,AU) field VLM systems have been discovered 
(Bill\'eres et al. 2005; Caballero et al. 2007; Artigau et al. 2007). These 
are not included in our analysis as they are not part of a defined sample. 
Nevertheless, their existence shows that at least some very wide VLM binaries 
could survive their birth environment.}

Weakly bound binaries can be disrupted by gravitational perturbations from both 
stellar encounters and giant molecular clouds, as discussed by Weinberg 
et al.\ (1987). Burgasser et al.\ (2003) applied the formalism of 
Weinberg et al.\ (1987) to the BD regime, and found that a pair of
0.05\,$M_{\odot}$ BDs separated by 10\,AU would be completely impervious to
disruption by stellar encounters in the field. Scaling the result of
Burgasser et al.\ (2003) to the most weakly bound system in our combined young 
VLMO sample, with a total mass of $0.16\,M_{\odot}$ and a separation of $100$\,AU, 
we get expected disruption timescales from field stars that are on the order 
of 600\,Gyr, more than 10$^5$ times older than the age of the systems.
These estimates are based on field star densities of $n_* = 0.05$\,pc$^{-3}$ 
(Bahcall \& Soneira 1980), which is a critical parameter, as the disruption 
timescale $\tau \propto n_*^{-1}$. But even 
assuming a stellar density as high as $n_* = 200$\,pc$^{-3}$, corresponding to 
the outer regions of the Orion Nebula Cluster (K\"ohler et al.\ 2006), the 
disruption timescale would be 150\,Myr, two orders of magnitude longer than the 
present age of the system. This estimate suggests that encounters with other 
stars do not disrupt even the most weakly bound systems in our sample, either in 
their birth environments (i.e., in OB or T associations) or later as the stellar
densities drop to match that in the field. Since perturbations from giant 
molecular clouds are insignificant in comparison to stellar encounters 
(Burgasser et al. 2003), the VLM binaries in our sample are likely to survive 
almost indefinitely.

\subsection{Frequency of Ultra Low Mass Companions}
In recent years, several VLM systems that appear to diverge from established trends 
have been reported. In particular very wide systems such as Oph162225-240515 with 
ultra low mass companions stand out from field VLM systems whose separation tends to 
decrease with mass. Because the presence of such systems imposes additional 
constraints on formation mechanisms, it is important to determine their frequency. 
Fig. 5 shows that our survey is sensitive to nearly planetary mass companions 
(down to 9M$_{\mathrm{J}}$) around 23 targets in the Cha I sample at separations 
greater than $100\mathrm{AU}$. In addition, such systems could be detected around 
all 34 targets from the Kraus et al. (2005, 2006) surveys of Upper Scorpius and Taurus. 
Yet, no ultra low mass companions at wide separations are detected in these three surveys, 
giving a frequency of no more than 2\% and thus confirming their rarity. 

Since the typical primary mass of the combined VLMO sample is 110M$_{\mathrm{J}}$, and 
the imaging surveys are sensitive to $q\sim0.1$ companions (i.e., secondaries with 
9--15M$_{\mathrm{J}}$) at $>100\mathrm{AU}$, we can also compare our results to surveys 
for sub-stellar companions to solar analogs. For example, Metchev \& Hillenbrand (2005) 
targeted 100 F5-K5 primaries to look for brown dwarf companions. Based on a single detection, 
they derive a sub-stellar companion freqency of at least $1\%$, and possibly on the order 
of a few percent. This result is in statistical agreement with the frequency of $q\sim0.1$ 
systems around VLM primaries in our combined young sample. Therefore, even though VLM systems 
with planetary mass companions are rare, the frequency of $q\sim0.1$ binaries in the VLM 
regime is no different from that for solar-type stars. 

\section{Concluding Remarks}
Our survey of VLMOs in Chamaeleon~I lends additional support to the established trend for a 
lower binary frequency with decreasing primary mass. By comparing the multiplicity 
properties of young VLMOs with those of field VLMOs and stars, we conclude that there is likely 
no evolution of multiplicity after the first few\,Myr for objects formed in low density regions.

While our survey of Cha I and the surveys of Kraus et al. (2005, 2006) and Konopacky et al. (2007) 
show that wide VLM systems with ultra-low-mass companions -- thought to be a critical challange to 
the ejection formation mechanism -- are infrequent, they also reveal a significant fraction of 
systems considered wide by the ejection scenario ($>20\mathrm{AU}$). Seven out of thirteen reported 
systems have separations wider than $20\mathrm{AU}$, with three systems separated by more than 
$\sim50\mathrm{AU}$. Bate et al. (2005) have recently reported that their simulations can produce 
binaries wider than the initial $20\mathrm{AU}$ limit, albeit via a different mechanism than the 
one responsible for the production of tighter pairs (i.e., simultaneous ejection and capture of 
previously unassociated objects vs. ejection of coupled tight binaries). However, since their 
hydrodynamic simulations are not followed up with N-body calculations, it is difficult to make a 
direct comparison; many of the reported simulated binaries are still accreting and likely unstable 
and the statistics are still somewhat limited; there may also be a spread in the ages of objects 
in a given SFR. In addition, most simulations model higher density regions, more akin to the Orion 
Nebula Cluster rather than the diffuse T associations that we and other observers have surveyed mostly. 
Nevertheless, a successful model for star formation has to account for observed binary properties in the 
VLM domain, including the existence of some wide pairs. 

\acknowledgments
This paper is based on observations collected at the European Southern 
Observatory, Chile under programs 075.C-0042 and 076.C-0579. We thank 
the ESO staff for carrying out some observations in service mode and 
for assistance during a visitor run. We are grateful to the anonymous 
referee for a careful review. This research was supported by an NSERC grant 
and an Ontario Early Researcher Award to RJ. MA was funded in part by an 
Ontario Graduate Scholarship and an NSERC Canadian Graduate Scholarship. 
DF received support from the Cosmo Grid project, funded by the Program for 
Research in Third Level Intitutions under the National Development Plan and 
with assistance from the European Regional Development Fund. 
This publication makes use of data products from the Two Micron All Sky Survey, which is a joint project of the University of Massachusetts and the Infrared Processing and Analysis Center/California Institute of Technology, funded by the National Aeronautics and Space Administration and the National Science Foundation.

\newpage

\begin{figure}
\begin{center}
\includegraphics[angle=0,scale=0.6]{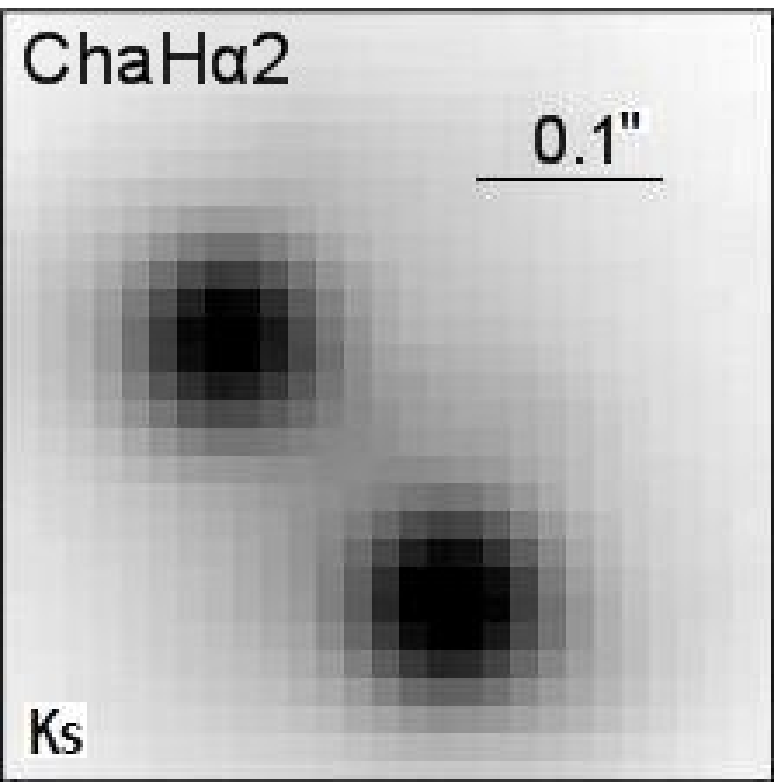} \ \ \includegraphics[angle=0,scale=0.6]{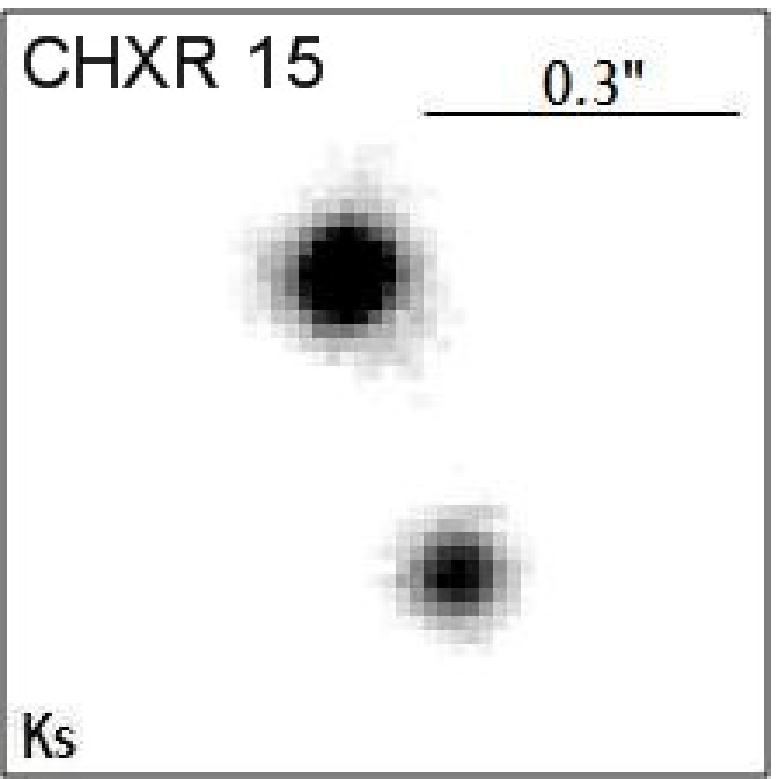} \ \ \includegraphics[angle=0,scale=0.6]{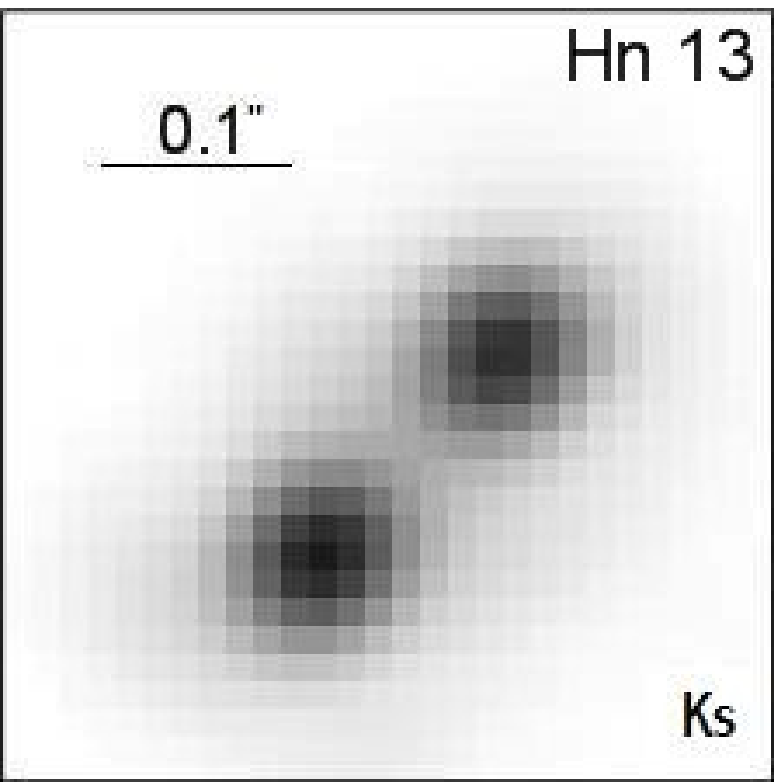} 
\end{center}
\caption{Resolved binary systems ChaH$\alpha2$, CHXR 15 and Hn 13 with separations of $0.164\pm0.003\farcs$, $0.305\pm0.002\farcs$ and $0.128\pm0.002\farcs$ respectively. All images are taken with $\mathrm{K}_{\mathrm{s}}$ filter. N is up and E is to the left in all three images.}
\label{Fig. 1}
\end{figure}

\begin{figure}
\includegraphics[angle=0,scale=0.8]{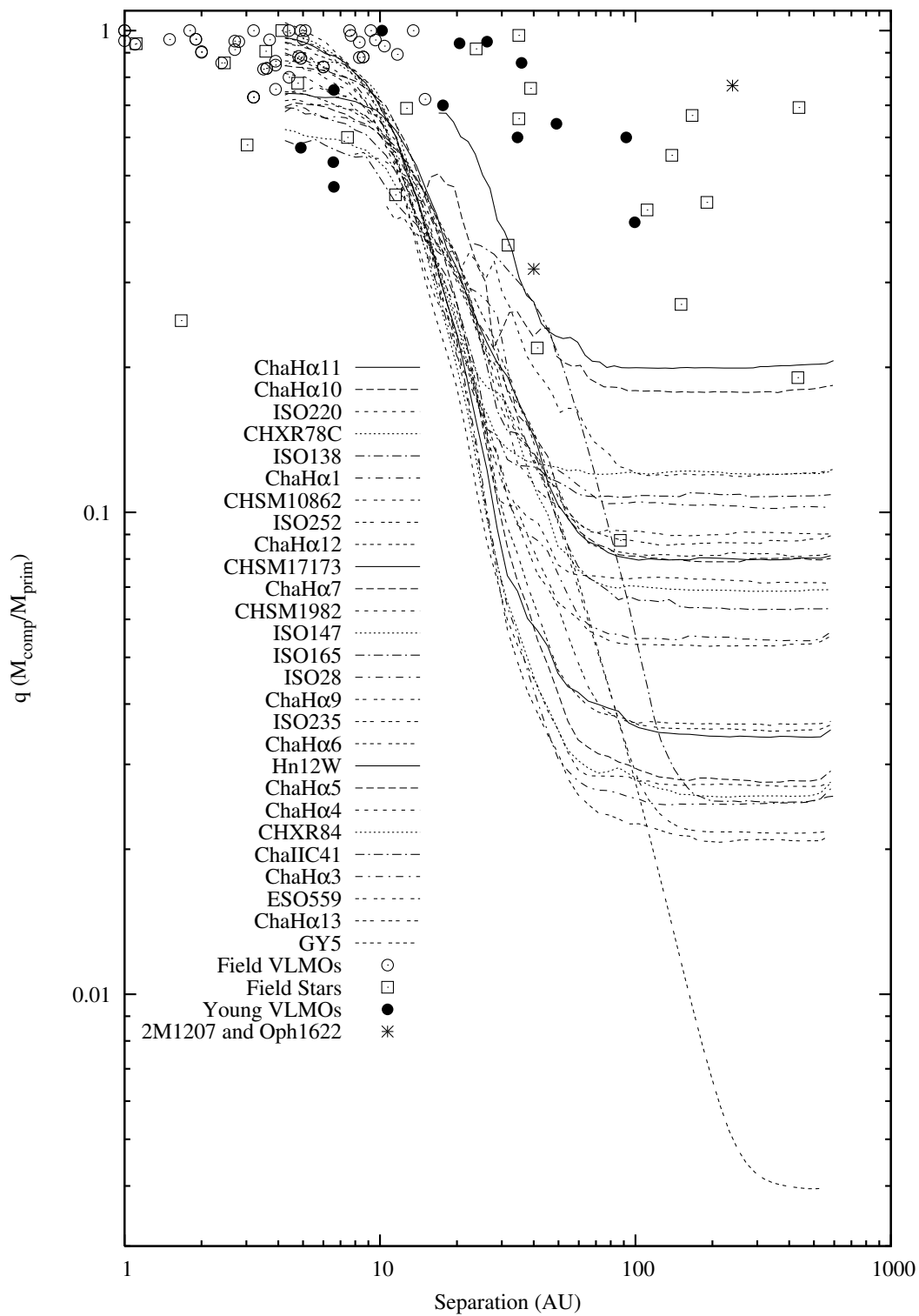}
\caption{Mass sensitivity plot for our survey targets (based on Baraffe et al. 1998, 2003 models). 
Also included are all resolved young VLMO binaries from Kraus et al. (2005,2006), Konopacky et al. (2007) as well as our Cha I sample, field VLMO binaries (Close et al. (2003), Burgasser et al. (2003), Siegler et al. (2005)) and field stellar binaries (Fischer \& Marcy (1992)). For a typical object, we are sensitive to Ks = 16.7 beyond 70AU. (best case Ks = 18.3, worst case Ks = 15.5)}
\label{Fig. 2}
\end{figure}

\begin{figure}
\includegraphics[angle=0,scale=0.8]{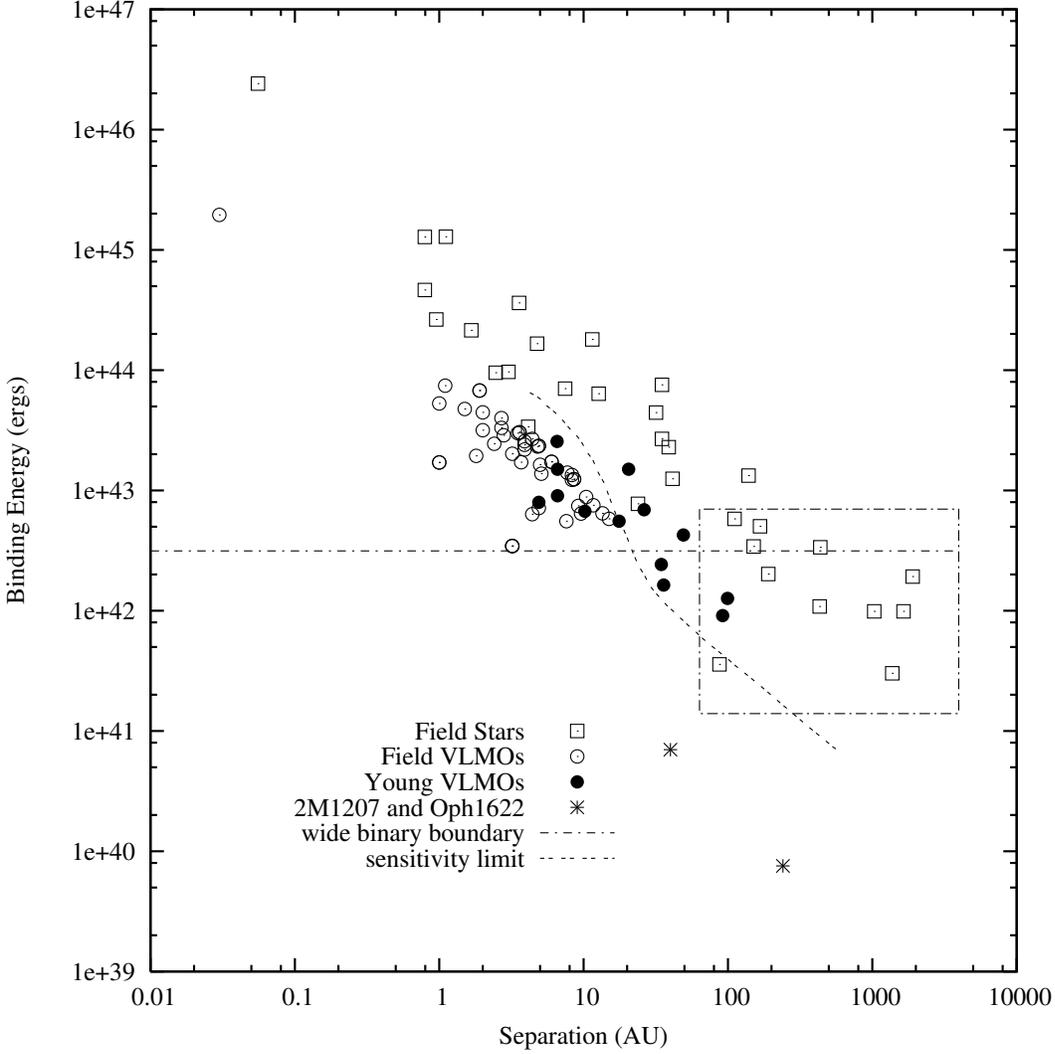}
\caption{Binding energy of young and field VLMOs and field stars. Symbols are the same as those in Fig. 2. We also include the mass dependant definition of wide systems explored by Burgasser et al. (2003), where a wide binary is any system such that $a>1400\times\left(\frac{\mathrm{M_{tot}}}{\msun}\right)^2\mathrm{AU}$. Four young VLMOs that are wider than the given limit come from the recent speckle survey of Taurus-Auriga by Konopacky et al. (2007); if these four are unresolved triples --i.e., one component is actually an unresolved equal-brightness binary-- then two of 
them would still be considered wide by the Burgasser et al. (2003) threshold, while if they are unresolved quadruples, all four would satisfy the derived limit. 
}\label{Fig. 3}
\end{figure}

\begin{figure}
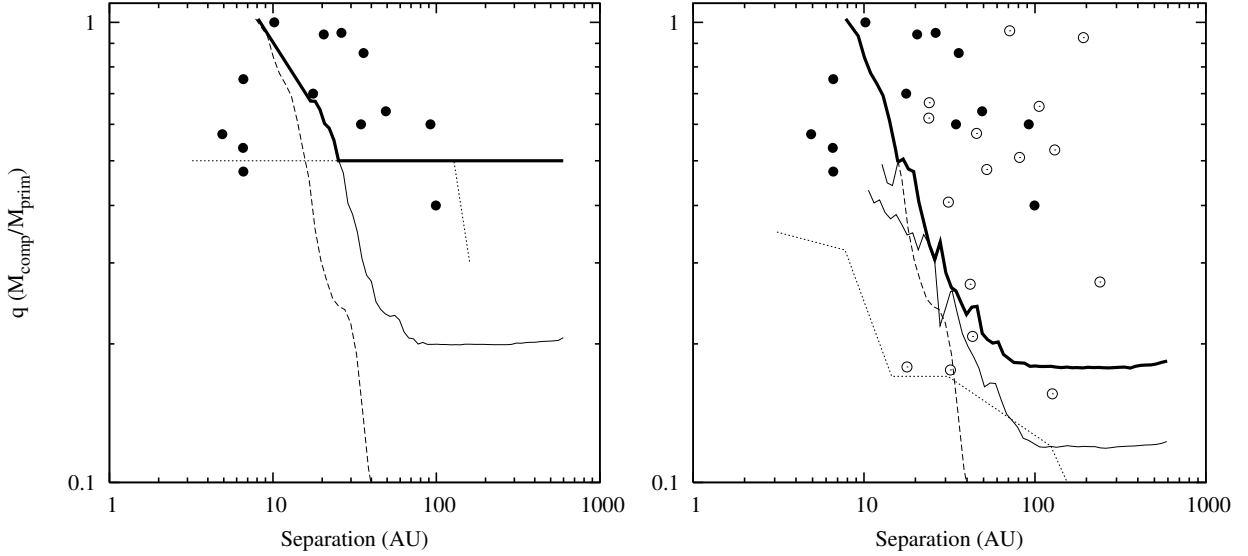

\begin{center}
\includegraphics[angle=0,scale=0.60]{f4a.epsi} \ \ \includegraphics[angle=0,scale=0.60]{f4b.epsi} 
\end{center}
\caption{Sensitivity limits used for comparisons -- (LEFT) between different mass bins in the combined young VLM sample -- (RIGHT) between the combined young VLM sample and young stars. We are sensitive to companions up and to the right of the thicker solid line which traces out sensitivity limits from the four surveys of young VLMOs (Kraus et al. (2005,2006), Konopacky et al. (2007) and this survey). Thin solid lines are the sensitivity limits from the Chamaeleon I sample, dotted lines are from the speckle imaging survey of Taurus (Konopacky et al. 2007), while dashed lines are sensitivity limits for the Kraus et al. (2006) Taurus sample, which were determined using the same techniques as those used for the Chamaeleon I sample. Since the Upper Scorpius observations are carried out with the same instument setup (ACS/HST), we assume the sensitivity limits for that sample are similar.
In the figure to the right, lower sensitivity limits are achieved by removing two worst offenders (one from the Cha I sample and one from Konopacky et al. (2007) Taurus sample) in order to include as many binaries as possible. The two offenders, especially the one from the Konopacky et al.\,(2007) survey, are atypical compared to the sample, justifying their removal in order to gain access to a wider parameter space. The solid dots are binaries from the combined young VLM sample while the open circles represent stellar binaries ($0.57-1.40\msun$) from Ghez et al. (1993,1997) surveys of Taurus and Chamaeleon I.}
\label{Fig. 4}
\end{figure}

\begin{figure}
\begin{center}
\includegraphics[angle=0,scale=0.6]{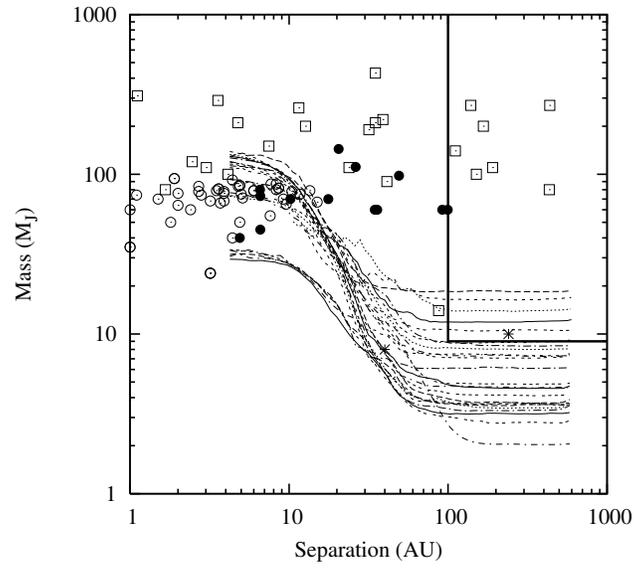}
\end{center}
\caption{Sensitivity limits of our sample in terms of mass. Once again symbols are the same as those in Fig. 2. The thick solid line outlines a region in the top right part of the plot in which we are sensitive down to 9M$_{\mathrm{J}}$ companions for most of our sample.}\label{Fig. 5}
\end{figure}

\newpage

\begin{deluxetable}{lccccccc}
\tabletypesize{\scriptsize}
\tablecaption{VLMOs imaged with NACO/VLT \label{TABLE 1}}
\tablewidth{0pt}
\tablehead{
\colhead{ID} & \colhead{$\alpha$(J2000.0)\tablenotemark{a}} & \colhead{$\delta$(J2000.0)\tablenotemark{a}} & \colhead{$K_S\tablenotemark{a}$} & \colhead{$SpT\tablenotemark{b}$} & \colhead{Mass($M_J$)\tablenotemark{c}} & \colhead{$T_\mathrm{eff}$\tablenotemark{d}}
} 
\startdata
ChaH$\alpha$11 & 11 08 29.27 & -77 39 19.8 & 13.54 & M7.25 & 60 & 2838\\
ChaH$\alpha$10 & 11 08 24.04 & -77 39 30.0 & 13.24 & M6.25 & 92 & 2962\\
ISO138 & 11 08 18.50 & -77 30 40.8 & 13.04 & M6.5 & 83 & 2935\\
CHSM17173 & 11 10 22.27 & -76 25 13.8 & 12.45 & M8 & 40 & 2710\\
ChaH$\alpha$7 & 11 07 37.76 & -77 35 30.8 & 12.42 & M7.75 & 45 & 2752\\
ChaH$\alpha$1 & 11 07 16.69 & -77 35 53.3 & 12.17 & M7.75 & 45 & 2752\\
CHSM10862 & 11 07 46.56 & -76 15 17.5 & 12.33 & M5.75 & 117 & 3024\\
ISO147 & 11 08 26.51 & -77 15 55.1 & 12.35 & M5.75 & 117 & 3024\\
CHSM1982 & 11 04 10.60 & -76 12 49.0 & 12.12 & M6 & 102 & 2990\\
ISO252 & 11 10 41.41 & -77 20 48.0 & 12.27 & M6 & 102 & 2990\\
ChaH$\alpha$12 & 11 06 38.00 & -77 43 09.1 & 11.81 & M6.5 & 83 & 2935\\
ISO220 & 11 09 53.37 & -77 28 36.6 & 12.23 & M5.75 & 117 & 3024\\
ISO28 & 11 03 41.87 & -77 26 52.0 & 11.69 & M5.5 & 134 & 3058\\
ESO559\tablenotemark{d} & 11 06 26.3 & -76 33 42 & 11.49 & M6 & 102 & 2990\\
ChaH$\alpha$9 & 11 07 18.61 & -77 32 51.7 & 11.80 & M5.5 & 134 & 3058\\
ISO165 & 11 08 54.97 & -76 32 41.1 & 11.44 & M5.5 & 134 & 3058\\
ChaH$\alpha$2\tablenotemark{d,e} & 11 07 42.45 & -77 33 59.4 & 10.68 & M5.25 & 153 & 3091\\
ChaH$\alpha$4 & 11 08 18.96 & -77 39 17.0 & 11.02 & M5.5 & 134 & 3058\\
ChaH$\alpha$3 & 11 07 52.26 & -77 36 57.0 & 11.10 & M5.5 & 134 & 3058\\
ChaH$\alpha$6 & 11 08 39.52 & -77 34 16.7 & 11.04 & M5.75 & 117 & 3024\\
Hn12W & 11 10 28.52 & -77 16 59.6 & 10.78 & M5.5 & 134 & 3058\\
CHXR84 & 11 12 03.27 & -76 37 03.4 & 10.78 & M5.5 & 134 & 3058\\
Hn13\tablenotemark{e} & 11 10 55.97 & -76 45 32.6 & 9.91 & M5.75 & 117 & 3024\\
ChaH$\alpha$13 & 11 08 17.03 & -77 44 11.8 & 10.67 & M5.5 & 134 & 3058\\
ChaH$\alpha$5 & 11 08 24.11 & -77 41 47.4 & 10.71 & M5.5 & 134 & 3058\\
ISO235 & 11 10 07.85 & -77 27 48.1 & 11.34 & M5.5 & 134 & 3058\\
CHXR78C & 11 08 54.22 & -77 32 11.6 & 11.22 & M5.25 & 153 & 3091\\
CHXR15\tablenotemark{e} & 11 05 43.00 & -77 26 51.8 & 10.23 & M5.25 & 153 & 3091\\
\hline
ChaIIC41\tablenotemark{d} & 12 59 09.8 & -76 51 04 & 11.4 & M5.5 & 109 & 3058\\
GY 5\tablenotemark{d} & 16 26 21.4 & -24 25 59 & 10.92 & M6 & 81 & 2990\\
2MASS1207-3932\tablenotemark{d} & 12 07 33.4 & -39 32 54 & 11.95 & M8 & 42 & 2710\\
\enddata
\tablenotetext{a}{2MASS Point Source Catalogue (Cutri et al. 2003), save for ESO559 and GY5, which are 
from Com\'eron et al. (2004) and Natta et al. (2002).} 
\tablenotetext{b}{Spectral types from Com\'eron et al. (2004) for ESO559, Natta et al. (2002) for GY 5, 
Barrado \& Jayawardhana (2004) for ChaII C41, Gizis (2002) for 2MASS1207-3932, 
and Luhman (2004) for all others. Effective temperature is obtained from the spectral types using the Luhman et al. (2003) temperature conversion scale. For the binaries, we assume that the
quoted effective temperature is the temperature of the
brighter component.}
\tablenotetext{c}{Masses calculated from $T_\mathrm{eff}$ and evolutionary models of Baraffe et al.\,(1998,2003) 
assuming an age of 2 Myr for Cha I, 1 Myr for $\rho$ Oph and Cha II, and 8 Myr for TW Hydrae. We also make a cautionary note: pre-main-sequence mass models are poorly calibrated and are also vulnerable to the age spread of Cha I, resulting in significant uncertainties.} 
\tablenotetext{d}{Object observed during the preliminary survey from May to September 2005.}
\tablenotetext{e}{Detected binaries. Masses of the primaries are listed.}
\end{deluxetable}

\begin{deluxetable}{ccccccc}
\tabletypesize{\footnotesize}
\tablecaption{Resolved Binaries \label{TABLE 2}}
\tablewidth{0pt}
\tablehead{
\colhead{ID} & \colhead{Sep($\arcsec$)\tablenotemark{b}} & \colhead{PA($\degr)$\tablenotemark{a,b}} & \colhead{Band} & \colhead{$m_A$\tablenotemark{b}} & \colhead{$m_B$} & \colhead{mass ratio\tablenotemark{c}}
}
\startdata
ChaH$\alpha$2 & $0.164 \pm 0.003$ & $40.6 \pm 1.9$ & Ks & $11.39 \pm 0.05$ & $11.47 \pm 0.06$ & 0.94  \\
Hn13 & $0.128 \pm 0.002$ & $318.0 \pm 1.9$ & Ks & $10.62 \pm 0.04$ & $10.70 \pm 0.04$  & 0.95 \\
Hn13 & $0.128 \pm 0.002$ & $318.7 \pm 1.9$ & H & $11.11 \pm 0.05$ & $11.25 \pm 0.05$  & --- \\
CHXR15 & $0.305 \pm 0.002$ & $200.0 \pm 1.4$ & Ks & $10.73 \pm 0.05$ & $11.33 \pm 0.07$  & 0.64 \\
CHXR15 & $0.304 \pm 0.003$ & $200.0 \pm 1.4$ & H & $11.14 \pm 0.05$ & $11.67 \pm 0.07$  & --- \\
\enddata
\tablenotetext{a}{Position angle is measured from north to east.}

\tablenotetext{b}{Photometry and astrometry measurements (along with uncertainties) of the two
components were obtained with the DAOPHOT routine in IRAF. Additionally, there
is a 1$\degr$ systematic uncertainty on the position angle. After measuring the flux
ratio of the primary and the secondary, we use it along with the 2MASS catalogue
K$_{\mathrm{s}}$ and H magnitudes for the combined flux to determine the magnitudes of each
component. We obtain the masses from effective temperatures.}  

\tablenotetext{c}{Uncertainties in the mass ratio are dominated by systematics (eg., 
unresolved higher order multiplicity, intrinsic variability of young VLMOs).}
\end{deluxetable}

\begin{deluxetable}{ccc}
\tabletypesize{\footnotesize}
\tablecaption{Statistical Sample Comparison \label{TABLE 3}}
\tablewidth{0pt}
\tablehead{
\colhead{\# of Binaries/Sample A Size} & 
\colhead{\# of Binaries/Sample B Size} & \colhead{Likelihood} 
}
\startdata
\sidehead{\textbf{Lower Mass Young VLMOs $(<0.07 \msun)$ \& Higher Mass Young VLMOs ($0.07-0.15 \msun$)}} 
0/27,0/21 & 8/45,6/39 & 0.02,0.08    \\
\hline
\sidehead{\textbf{Young VLMOs ($0.04-0.15 \msun$) \& Young Stars ($0.57-1.40 \msun$)}}
9/70,7/58 & 12/34,12/34 & 0.02,0.01    \\
\hline

\enddata
\tablecomments{Entries listed in each column represent samples and statistical comparisons with and without Upper Scorpius members.}
\end{deluxetable}  
\end{document}